\documentclass[12pt,preprint]{aastex}
\slugcomment{Accepted for publication 
in {\it the ApJ}  August 1, 2003} 

\journalid{ApJ}{}
\articleid{}{}

\begin{document}
\def\lax    {\ifmmode{_<\atop^{\sim}}\else{${_<\atop^{\sim}}$}\fi}
\def\gax    {\ifmmode{_>\atop^{\sim}}\else{${_>\atop^{\sim}}$}\fi}
\def\gtorder{\mathrel{\raise.3ex\hbox{$>$}\mkern-14mu
             \lower0.6ex\hbox{$\sim$}}}
\def\ltorder{\mathrel{\raise.3ex\hbox{$<$}\mkern-14mu
             \lower0.6ex\hbox{$\sim$}}}

\title{A Method for Black Hole Mass Determination in 
Accretion Powered X-Ray Sources}

\author{Chris R. Shrader$^{1,3}$,  \& Lev Titarchuk$^{1,2,4}$}

\altaffiltext{1}{Laboratory for High Energy Astrophysics,
NASA Goddard Space Flight Center, Greenbelt, MD 20771, USA;
Chris.R.Shrader@gsfc.nasa.gov,titarchuk@lheapop.gsfc.nasa.gov}
\altaffiltext{2}{E.O. Hulbert Center for Space Research, Naval Research 
Laboratory}
\altaffiltext{3}{Universities Space Research Association, Lanham MD}
\altaffiltext{4}{George Mason University/CEOSR, Fairfax VA}
% check NRL affiliation?

\rm

\vspace{0.1in}

\begin{abstract}
We describe a method for the determination of black-hole masses based on
information inferred from high-energy spectra. It is required that the spectral
energy distribution consist of thermal and 
%non-saturated 
Comptonized components. One can then, in principle, infer the depth 
of the gravitational potential well for sources of known distance.  The
thermal component is inferred by the integration of a
blackbody spectral form over the
disk. We assume that the color temperature distribution in the disk  has a
specific shape given by the Shakura-Sunyaev (1973) disk model  which goes to
zero  at the inner disk radius and at infinity and has a maximum at 4.2 
Schwarzchild radii. In this formulation there is
only one parameter, the so called color correction 
factor, relating the apparent
temperature to effective temperature which characterizes the thermal emission
component. We have made use of improved Galactic black hole binary
dynamical mass determinations to derive, in 
effect, an empirical calibration of
this factor. We then present our analysis of observational data for
representative objects of several classes;  
Galactic black hole X-ray binaries, narrow line
Seyfert galaxies (NLS1), and ``ultra-luminous'' 
extragalactic X-ray sources (ULX). We then
apply our mass determination calculation and present our results.  
We argue that this approach can
potentially fill a void in the current knowledge of NLS1 and ULX properties,
and discuss how a deeper understanding of both classes has relevance to the broader
issues of how cosmic black holes, beyond the stellar-mass realm, are formed 
and what is their overall mass distribution.
\end{abstract}

\section{Introduction}

Recent developments leading to renewed interest in
the problem of the cosmic distribution of black hole masses, their origin  and
growth, have motivated us to propose a new method for black hole mass 
estimation. In galaxies, the
bulge $M-\sigma$ relation, its apparent cross-calibration with reverberation 
mapping methods and related correlations have provided a greatly expanded 
database of central black hole masses, and support 
the idea that masses
$\sim(10^6-10^7)M_{\odot}$ are a  ubiquitous feature of galaxies 
(e.g. Ferrarese \& Merrit 2000; Wandel, Peterson \& Malkan 1999; Woo \& Urry 2002).
However, velocity dispersion data are difficult to obtain 
for a large sample of objects, the reverberation mapping efforts are
observationally intensive to compile and the correlation methods potentially
suffer from calibration uncertainty.
In a number of external galaxies, notably ones with active
start  formation regions, the so called "Ultra-luminous" X-ray sources (ULXs)
have led to  speculation on the existence of intermediate ($10^2 - 10^4)
M_{\odot}$ mass objects [e.g. Colbert \& Mushotzky 1999; Strohmayer
\& Mushotzky 2003; Miller et al 2003). However, the prospects for 
dynamical measurement supporting these assertions are poor.

In parallel with these observational developments,  a growing inventory of
dynamically determined Galactic black  hole masses in accretion powered
X-ray binaries (herein BHXRBs) continues to emerge (e.g. Orosz 2002). We
would thus suggest that pursuit of an  independent methodology for BH
mass determination is desirable, in particular, one which can be calibrated
through use of this knowledge base. The method we propose makes use of
the high-energy spectra  of these objects. 
Our work builds upon Borozdin et al. (1999)
hereafter BOR99, and  Shrader \& Titarchuk (1999) 
hereafter ShT99, where we introduced 
our basic methodology [also refer to Colbert \& Mushotzky (1999), 
in which one of us (LT), contributed calculations leading to one of 
the earliest claims of intermediate black hole (IMBH) existence]. 
The major differences here are 
that the a greater number of the Galactic binaries  
are dynamically constrained and can thus be used to test
our results, and that we have applied the analysis to 
representative objects of the ULX and NLS1 classes. We also present some 
additional computational details and expanded discussion that were 
absent in the earlier papers.

We further explore possible links to studies of the
aforementioned extra-galactic populations. The basis for this extrapolation is
the similarity between the X-ray spectra of a subclass of active galaxies, the
narrow-lined Seyferts (NLS1s) and the galactic black-hole binaries. Indeed,
it is now apparent that at least some ULXs most likely 
are accretion driven compact objects (Strohmayer \& Mushotzky 2003), 
and in at least a few cases exhibit the well known bi-modal
spectral behavior seen in galactic objects. 

The emergent X-ray spectrum of an accretion driven compact source is 
generally a convolution of a thermal source component 
associated with an  accretion disk 
flow  and the scattering Green's function which represents the effects of  
Comptonization by energetic  electrons. A deconvolution of these two
effects, in principle, allows one to derive with minimal model dependence a
physical normalization term for the thermal component from which a
distance-to-mass ratio can then be determined. This is the basic idea
underlying our method.  One particular spectral model is the bulk-motion
Comptonization  (BMC) model (Titarchuk, Mastichiadis \& Kylafis 1996,
1997; Laurent \& Titarchuk 1999, 2001),  which  has already been applied to
this problem by Shrader \& Titarchuk (1998), hereafter ShT98 and ShT99.   However, most of what we discuss here is independent
of the  particular geometrical  and  dynamical configuration of the
Comptonizing medium.  The minimal assumption is required in 
constructing the multicolor disk spectrum is that the
release of gravitational energy of the matter in the disk, presumably due
to viscosity, diminishes toward the inner disk boundary.  Novikov \&
Thorne (1973) were the first to formulate this boundary condition [see also
Shakura \& Sunyaev (1973), hereafter SS73, for detail  of the disk structure]. 
Combination of the viscous dissipation of the gravitational energy and this 
boundary condition leads to the formation of the color
temperature distribution over the disk with a characteristic maximum
at about 4.2  Schwarzchild radii (SS73). We note that we do not consider
the case of extreme Kerr geometries; further discussion of this issue is
presented in section 4.
 
 We assume that the color temperature distribution in the disk has a specific
shape, given by the SS73 disk model, which goes to zero at the inner
disk radius and at infinity and has a maximum at $R_{max}=4.2 R_{\rm S}$ 
where $R_{\rm S}=2GM/c^2$ is Schwarzchild radius. This formulation
of the disk spectral model, obtained by  integration of blackbody spectra over
the disk, has only one parameter; the so called color correction 
factor $T_h$. The color correction factor relates
the apparent temperature $T=T_{col}$ to effective temperature $T_{eff}$ 
which characterizes the
thermal emission component. The color factor $T_h=2.6$ was calibrated 
using the spectral measurements of GRO J1655-40 for which the distance
and  mass of the central objects are known accurately 
(BOR99). 

We note that BOR99 also estimated the central object mass in GRS
1915+105,  $m=M/M_{\odot}=(13.5\pm 1.5)/(\cos{i})^{1/2}$ with an assumption
that the same  color factor $T_h=2.6$ is  valid for the disk in this source.
Recent dynamical measurements of Greiner et al. (2000), who find a mass of
$14\pm 4$ in GRS 1915+105, further corroborate our results. We were thus 
encouraged to undertake an effort to expand our study of 
galactic binaries, to further
calibrate and, hopefully, validate our method. We then apply the same
methods to selected objects among the aforementioned extra-
galactic classes  for which the high/soft state spectral form is observed. 
   
In this paper, we present the details of our mass-determination method 
(\S~2), and then proceed to apply our analysis to observational data
(\S~3). This includes samples of additional galactic X-ray binaries, 
NLS1s and several  ultra-luminous X-ray sources in nearby galaxies.
Comparison to independent mass determinations, and the implications of
some of our results are discussed in \S~3 and we summarize and draw
conclusions in \S~4.

\section{Description of Mass Determination Method}

To summarize the idea of our method we  enumerate the main points:
(i) We extract as accurately as possible the soft component from the data 
using the exact form of the Comptonized spectrum derived in Sunyaev \& 
Titarchuk (1980); Titarchuk (1994), Titarchuk \& Zannias (1998).
(ii) We calibrated the color hardening factor $T_h=T/T_{eff}$ using the GRO 
J1655-40 for which the distance and the mass are known. We argue that this 
approach is superior to model-dependent methods (e.g. Shimura \& Takahara 1995) 
given the number of unconstrained free parameters. We note that Merloni 
 Fabian \& Ross (2000) and also Hubeny et al. (2001) demonstrate the wide 
range of possible model-specific $T_h$ values.
(iii) When $T_h$ is established, one can proceed with mass determination 
using the color temperature $T$ and the absolute normalization of the 
soft component $A_n$ which depends on the mass, distance, inclination 
angle {\it i} and the effective emission area radius $r_{eff}$ that in turn 
depends on $T$ only (see BOR99, Eqs 4 and 6) allowing us to infer 
distance-to-mass ratio with an accuracy of $cos^{1/2}i$.
(iv) Our model (using SS73, see details in BOR99) then allows for determination 
of the effective area,  from the observationally determined color temperature $T$.

In Figure 1 we present an artistic  conception of our model.
The low-energy blackbody X-ray photons associated with
accretion powered objects such as BHXRBs, NLS1s or ULXs,  are assumed  to come 
directly from the thin accretion disk, as is well established to be the case in
BHXRBs. The higher energy photons that form the  power law tail of the
spectrum come from upscattering (acceleration) by energetic electrons  
rushing towards the central black hole. The  plasma dynamics is a mixture of 
thermal and bulk motion inflow. Some part of the disk soft  photon flux comes directly to the Earth observer
but some fraction of the disk flux $Q_d$ comes after the scattering in the
corona region (the site of the energetic particles).  In fact, the corona could be
very hot if the disk illumination flux $Q_d$ is much less than the energy
release in the corona $Q_{cor}$ and it can be relatively cold if 
$Q_d\gtorder Q_{cor}$ (see more details of these picture  in Chakrabarti
\& Titarchuk 1995). 
% analogous to the shock acceleration of particles. 

To determine the central object mass, the basic idea is to 
extract the blackbody component from
the emergent spectrum whose shape is modified by Compton scattering off the
hot electrons.  The intensity of the blackbody radiation we measure depends
on the emission area of the accretion disk and the distance to the source,
which for extragalactic objects is usually known with reasonable precision.
Depending on the color temperature, typically $kT \simeq 1$~keV for
BHXRBs, the effective radius associated with this area can be 
5-15 Schwarzchild radii. 

The next problem is to determine a ratio of color to effective temperature,
the so-called color factor, $T_h$. This characterizes the deviation between
the emergent spectrum and a pure blackbody. For the pure blackbody case,
the luminosity per unit area is a product of the fourth power of the
temperature and the Stefan-Boltzmann constant; this is easily generalized to
the multi-temperature disk scenarios. Equating the observed, integrated
luminosity  to this expression yields a different temperature from  that
inferred from a spectral fitting procedure using a Planck distribution. This
phenomenon is well known in solar  and stellar  astrophysics. For example,
the Sun has an effective temperature of 5500 K,  but its spectrum is not
described by a 5500-K black body. Physically, this difference  results from
scattering effects in the solar atmosphere, through which we view the 
substrata to about an optical depth. In the case of galactic BHXRBs, the
appropriate color  factor can be determined theoretically by invoking several
assumptions regarding the  accretion disk, or it can be empirically calibrated
using a source for which the black hole  mass and distance are already 
known through independent methods. One
must then assume that the color factor inferred is invariant to a
reasonable approximation (see more discussion of this issue in \S 3).

As noted, BOR99 previously calibrated the color factor using a particular
source,  GRO~J1655-40, for which accurate mass, distance and orbital
inclination measurements have been made (Orosz \& Bailyn 1997; Orosz 2002). 
The  color factor
thus obtained was $T_h \simeq 2.6$. We can now expand upon that effort,
making use of the additional dynamical binary solutions which have emerged.
Once established, $T_h$ can be used along with inferred spectral parameters
in a distance-to-mass determination. For Galactic sources, distances are often
crudely  determined, but in external galaxies they are more precise. A major
question which we must address however, is whether or not the color factor
is similar for other classes of objects, which are in different environments
and different regimes of $M$ and $\dot M$.

\subsection{Color Factor Calibration}

 We have made use of a growing database of dynamically constrained
black-hole masses in  Galactic binaries (e.g. Orosz 2002) 
to better calibrate this key parameter
of our method; the color correction or "hardening" factor.  In BOR99 and
ShT99 details of the BH mass determination procedure
were presented. For completeness, we duplicate some description
of that procedure for the determination of the $m/d-$ratio here, augmented with
some additional details, and arguments supporting  its generalization to other 
object classes. 

The inferred absolute normalization of the disk multicolor disk,
presented in approximation of as a single blackbody spectrum reads as
follows
\begin{equation}
A_N=C_N
r_{eff}^2=\frac{0.91m^2\cos{i}T_h^{-4}}{d^2}r_{eff}^2(T_{col}),
\end{equation}
where $d$ is the distance to source in kiloparsecs, $\cos{i}$ is the cosine of
the inclination angle (the angle between the line of sight and the normal to
the disk) and $r_{eff}(T_{col})$ is  the effective radius of the area related to
the best-fit blackbody color temperature $T_{col}$.   The blackbody spectral
shape with color temperature $T_{col}$ provides a reasonable
approximation  to the multicolor disk spectrum (e.g. BOR99, eq. [2]) if the
ratio of energies (for a given photon energy band) to the best-fit color
temperature is greater than 2 (see BOR99, Fig.4). It is worth pointing out 
the values of the best-fit parameter $T_{col}$ of the observed spectrum and  our
best-fit parameters  $r_{eff}$ and $T_{col}$ are sensitive to the 
energy range of a given instrument. 
One should be very careful to take this into account when applying 
our technique (see details in BOR99 and ShT99)

In the Bulk Motion Comptonization model (BMC) the absolute
normalization of the blackbody component 
$\tilde A_{N}$  is related to $A_{N}$:
\begin{equation}
A_N=\frac{8.0525 \tilde A_N^{bb}}{(1+f)[kT/(1{\rm keV})]^4}
\end{equation} 
where $f$ is the so-called illumination parameter, which for the
Comptonized component is 
\begin{equation}
A_{N}^{comp}= A_N^{bb}f/(1+f).
\end{equation}
In a case of $f\ll 1$ 
the Compton cloud (presumably in the high/soft state it is the converging inflow)
 occupies relatively small area, characterized
by typically less than 3
Schwarzchild radii and a predominantly thermal spectrum is observed. 
By comparison, the  effective disk area, characterized by
$r_{eff}=R_{eff}/R_{\rm S}$, was 
found to be more than $10$ by Borozdin
et al. (1999). 

 There are cases when $f\gg1$, for example, during the
low/hard state. This indicates that the Compton cloud  completely covers the
effective disk area. 
%Because scattering in the Compton cloud does not
%significantly boost a large fraction of the soft disk photons,
 In this case we can also
determine $A_N$ using eq. (2). To do this one can also use $\tilde
A_N^{bb}$ because $\tilde A_N^{comp}=\tilde A_N^{bb}f/(1+f)\sim \tilde A_N^{bb}$. 

In BOR99 we inferred the hardening
factor $T_h=2.6\pm0.1$ for GRO J1655-40 using formula (1) in which there
was only one  unknown parameter, $T_h$. Specifically, we use $m=7\pm 1$,
$i=70^o$ (Orosz \& Bailyn 1997),  $d=3.2\pm0.2$ (Hjellming \& Rupen
1995). The effective radius $r_{eff}$ as a function of the best-fit color 
temperature was  calculated (for RXTE data see Fig. 5 in BOR99 and formulas
in ShT99). The inferred coefficient $A_N$ was
obtained using the best-fit parameters of the BMC model, namely $\tilde
A_N^{bb}$, $T_{col}$ and $f$ (see formula 2).

\section{Application to Observational Data}

We obtained data from the NASA  High Energy Astrophysics Science
Research Center (HEASARC) for representative samples 
of BHXRBs, NLS1s, and a few
ULX sources. The database utilized for the BHXRB analysis was primarily
derived from the the Rossi X-ray Timing Explorer (RXTE) archive, although
in several cases this was supplemented with higher energy data from the
Compton Gamma Ray Observatory.  We used archival data from the ASCA
X-ray satellite for most of the NLS1s in our sample, supplemented by
publicly available Chandra data for the "type-I" quasar PHL~1811. For the
ULXs, we used our spectroscopic analysis was based on ASCA CCD
and gas-scintillator spectra (NGC 1313, NGC 5408 and M33), and on
Chandra ACIS CCD spectra (NGC 253, NGC 1399 X-2 and X-4).
In addition, we examined Chandra images of the ASCA fields
to assess possible source confusion
situations. We also obtained and analyzed archival XMM/MOS data for 
the NGC~1313 source, and Chandra and XMM archive data for the M82 ultraluminous 
source. In the case of NGC~1313, the observations 
with XMM were made during an apparent low-hard state of that source, thus 
making it difficult to deconvolve the thermal and Comptonized components,
so we did  not use these data in our final analysis. The results presented
for that source are based on one of two archived ASCA observations, 
when a high-soft-state spectral form was exhibited. Similarly, we were unable
to interpret the spectrum that we derived for the M82 source in the context
of our model (or any obvious thermal plus Comptonization model, although see
Strohmayer \& Mushotzky 2003), thus we 
did not include that object. 

We performed forward-folding spectral analysis applying the bulk-motion
Comptonization model to obtain the normalization parameter described in
Section 2, and in BOR99 and ShT99. The similarity between 
the BHXRBs, NLS1s and ULXs is
evidenced in Figure 2; here familiar HSS BHXRB spectral form is mimicked
by the other representative objects (although the NLS1 characteristic
temperature is much lower). 

\subsection{Galactic BHXRBs}

 We have applied our analysis to a sample of 13 Galactic BHXRBs, as 
indicated in Table 1; also see to Figure 3. 
As noted, we have used GRO~J1655-40 as our
primary calibrator, with additional self-consistency checks such 
as GRS~1915+105, XTE J1859+226 and XTE~J1550-56. The accuracy of 
the distance estimates vary considerably over our sample, and
this is a  major component to the uncertainty in our mass determination; 
as noted we are really computing the mass-to-distance ratio. We have 
thus included a distance column in Table 1 with numbers drawn from
the recent literature, indicating the value used in our calculation.

The most notable discrepancy between our results, and those tabulated in
Orosz (2002; also see Shahbaz, Naylor \& Charles 1997) 
for Nova Muscae 1991 and LMC X-1. For Nova Muscae 1991,
our mass estimate of $12.5 \pm 1.6$ exceeds the dynamical mass by
about $2.5\sigma$. Also for LMC X-1, which is poorly constrained 
dynamically, our result is about $1\sigma$ above the upper bound 
in Orosz (2002). 

We made a self-consistency check by estimating 
the mass accretion rate in the disk $\dot m_{disk}$ in cases
where the the distance and thus mass are known.
We remind the reader that $\dot m_{disk}=\dot M/\dot M_{\rm E}$
where $\dot M_{\rm E}=L_{\rm E}/c^2$ and $L_{\rm E}=4\pi GMm_p/\sigma_{\rm T}c$. 
We then applied the formula for $\dot m_{disk}$  inferred in BOR99
(see eq. 8 of that reference). We note that Chakrabarti \& Titarchuk (1995) 
argued that $\dot m_{disk}$ should be of order of unity or higher
in the high/soft state. One can also calculate the luminosity 
of the soft component, $L_s$, 
using formula (9) in BOR99 once $m$ and $\dot m$ are determined.
{\it For all Galactic and extragalactic  BH sources that we have
analyzed, we find that the high-soft condition 
$\dot m_{disk}\gtorder1$ is satisfied and $L_s<L_{\rm E}$}
(in most cases $L_s$ is a few percent of $L_{\rm E}$).

\subsection{Narrow-Line Seyfert-1 Galaxies}

 The narrow-line Seyfert-1 Galactic Nuclei (NLS1s) are another class of
objects to which our  technique can be applied. These objects, summarized in 
detail in elsewhere (e.g. Leighly 1999a; 1999b; Vaughan et al 1999), are
found to have an excess of soft X-ray emission relative to a powerlaw
component, which is in turn softer (photon indices 
$\Gamma \simeq 2-2.5$, as opposed to
$1.7-1.8$) than that of the broader class of 
Seyfert-1 galactic nuclei. As has been noted, their spectra often
resemble those of Galactic black-hole X-ray  binaries in the 
high-soft state (Figure 2).
Implicit in our analysis is the assumption that  the soft-excess, as in the
Galactic binaries, associated with high mass-accretion rates. 

There have been
suggestions that the masses of the central compact objects in the NLS1s may
be systematically lower, perhaps by as much as 
a factor $\sim10$ than those of typical broad-line Seyfert  galactic nuclei. 
One possible explanation is that the black holes in these systems are still
being grown, and thus the copious soft X-rays. However, this may be an overly
simplistic an argument, and the empirical picture needs to be further solidified.

In a number of cases, independent estimates  are available from reverberation 
mapping  and/or bulge mass/velocity dispersion correlations  (e.g. Wandel,
Peterson \& Malkan 1999)   methods, so a direct cross-comparison  of our
results  to those methods can be made. More recently, cross-calibration of
those methods  with observational properties which are more easily obtained, 
such as bulge magnitude or narrow emission-line properties. 
This has led to much more extensive
compilations (e.g. Woo \& Urry 2002). From  these  tabulation in the
literature, we can compare our results as a cross check,  and to see if any
systematic trends are evident.

To further explore these issues, we have revisited observational data for   a
subset of the NLS1 population. We have applied our fitting  procedure, as
described above, to the NLS1 sample listed in Table 2. From the parameters 
enumerated in Table 2, using the luminosity distances determined form the
known source  redshifts, we can then estimate the central BH mass following
the procedures outlined in section 2.

Generally, the functional form of the model fits resemble the Galactic
black-hole X-ray  binary high-soft-state. The characteristic blackbody
temperatures however, are typically  lower, $kT \simeq 0.1$ keV. We have
studied the issue of the color factor determination for these lower
temperature environments. 
This issue has been previously addressed in
radiative transfer calculations
in accretion disks of super-massive objects (Hubeny et al.
2001, hereafter HBKA). The analytic and numerical radiative transfer models
in HBKA are reasonably self consistent and are, in principle, able to
produce high-temperature external layers, the so called ``disk corona''. 
Comptonization in these corona leads to an increase in the color factor $T_h$ 
dependent upon a model parameter of the disk, $\epsilon$ known as the 'mean photon
destruction parameter". $T_h$ can vary from 2 to 40 as $\epsilon$ changes
from $10^{-2}$ to $10^{-8}$.  Such a wide range possible of $T_h$ values
does not allow us to specify a certain value of $T_h$ using the HBKA disk
model but it demonstrates that high values of $T_h$ may not be
excluded. We found that the our set of mass estimates for supermassive
objects would be agree reasonably well with those derived from other methods
if $T_h \simeq 12$.  Our results are depicted graphically in Figure 4.

\subsection{ULXs}

Another class of objects which have been seen to exhibit the high-soft/low-
hard state bimodality are the ultra-luminous X-rays sources (ULXs) seen in a
number of nearby galaxies, notably in or near star forming regions (e.g. Ptak
\& Colbert 2002). Ultra-luminous in this context refers to apparent
luminosities  in the $\simeq 1-10-$keV X-ray band of $\sim1.5\times10^{38}$ erg
s$^{-1}$, which is well  in excess of the Eddington limit for solar mass
objects (Colbert \& Mushotzky (1999); Griffiths et al. 2000; Kaaret et al. 2001).  Furthermore, the
spatial resolution of the Chandra X-ray  Observatory has demonstrated that
in a number of  cases, such objects are clearly distinct from the
dynamical center of the host Galaxy.  This in particular has generated
considerable interest, as one possible explanation which avoids the
Eddington-limit problem is
that objects of  "intermediate" mass scale, $\simeq 10^2 - 10^4$
M$_{\odot}$ are present in interacting binaries. Alternative hypotheses,
invoking beaming have also been suggested to avoid this problem (King et al
2001), but this may be problematic, as at least some  objects apparently
illuminate reflection nebulae. Also, the spectral energy distribution in at
least some cases has a distinctly thermal component, and perhaps most 
significantly, the recent discovery of quasi-periodic oscillations in the M82 
source argue very strongly for a compact disk-fed system,
quite likely a binary (Strohmayer \& Mushotzky 2003). We thus suggest that this
is a pertinent area of investigation for alternative mass-estimation methods.
We note that this has already been calculated using the ShT99 technique
in a preliminary manner and presented in Colbert \& Mushotzky (1999) .

We should also note that there are instances of Galactic or Megallanic 
cloud X-ray binaries which have exhibited luminosities overlapping the lower
extent of the nominal ULX luminosity distribution (e.g. McClintock \&
Remillard 2003). We attempt to address this in section 4.
    
    We have applied our model fitting and mass determination procedure to 
7 objects in 6 galaxies categorized as ULXs. Each are previously noted in the
literature to exhibit an apparent thermal excess
plus unsaturated Comptonization spectrum; (NGC~1313, NGC~5408 \& M33:
Colbert \& Mushotzky 1999; NGC~253, NGC~1399 X-2, NGC~1399 X-4; Humphrey
at al 2003). Results are presented in Table 
3 (also see Figure 6). The most compelling cases for the presence of 
intermediate mass objects are for the NGC~1313 and
NGC~5408 objects. Our calculations for the NGC~253 source, and 
the two NGC~1399 sources also suggest objects of intermediate mass
in the $10^{3}-10^{4}$ range, however, we consider these results less
certain.  These sources are faint, and
the spectra we were able to extract consist of only 500-1000 net
photons. This is reflected in the size of the error bars associated 
with those sources in Figure 5, which is ultimately due to our inability to 
better constrain the $kT$ and normalization parameters of our spectral
deconvolution. 

Our results the M33 source however, are consistent with it
being a few $\sim 10$ solar mass object, perhaps not entirely dissimilar
from GRS~1915+105. As noted, we were unable to obtain useful results in 
the context of our methodology for the M82 source, or for the apparent
low-hard-state situations in the NGC~1313 observations of July 12, 1993 (ASCA)
or October 17, 2000 (XMM).  The shortcoming in each of these instances
was our inability to reliably extract a thermal component.

\section{Discussion and Conclusions}    

To summarize: (i) we have presented a method for the
determination of central black-hole masses based on X-ray spectral
extraction (ii) the principal parameter of the method, the X-ray color factor
has been established empirically for BHXRBs and NLS1 subclasses (iii) the
NLS1 mass distribution we find is offset from the distribution of normal
Seyfert-1 mass estimates. In small overlap between our sample and published
mass estimates from reverberation mapping, there is good
agreement in one case (Mkn 110), poor agreement in ones case (NGC 4051),
and nominal agreement in another (Mkn 335). (iv) The results we obtain for
NGC~1313, NGC~5408, NGC~253, NGC~1399  are consistent 
with the existence of intermediate mass
objects in those systems, specifically black home masses of $\sim 170$,
$\sim 110$, $\sim 10^3$ and $3.5\times 10^3$ solar masses 
respectively. For the M33 source, the our mass estimate 
suggests an $\sim 30$ solar mass object, so it may be an extreme
manifestation of a supernova produced stellar mass black hole.
(v.) Our inferred estimates of luminosity using the best-fit 
parameter $T_{col}$, and inferred BH masses [see equation (9) 
in BOR99, for details] show that for the most of the analyzed sources the
bolometric luminosity in the high/soft state is a few percents 
of the critical (Eddington) luminosity.
 
There are some instances in the literature where other
authors have cited difficulty in reconciling the observed 
temperatures with intermediate BH masses (see e.g. Kubota et al.
2002). We revisited some of those same data within the context 
of our own analysis, notably for the IC 342 "source 1",
 which was the basis of Kubota et al. (2002) paper.
%The Kubota et al. paper has, in our view, a number of problems, 
%mainly in their what we view as an over-interpretation of the Tin-Ldisk 
%diagram. The temperature of the inner disk radius is not observable 
%quantity, rather it is a inferred parameter of their particular disk 
%model. Their model is an approximation of the SS73 that does not take 
%into proper account the inner disk boundary condition. The contribution 
%to the spectrum of the innermost region is thus overestimated. A 
%self-consistent analysis of the disk emission area was established in 
%BOR99 and ST99, revealing that it exceeds the inner-most area by an 
%order of magnitude. In practical terms, this means that there is little 
%spectral information on this region.
%Furthermore, the high frequency QPO (related to the rotational Kepler frequency
%the innermost disk area) is not seen as oscillations of the soft component but
%it is rather seen as the oscillations of the power-law component of the
%spectrum (see e.g. McClintock & Remillard 2003).  Thus, the 
%difficulties troubles  that group has with interpretation 
%of the ULX (or BHXRB) physical parameters results from their 
%approximation of the SS-disk model. They 
%also invoke a rough approximation of the Comptonization spectrum.  
We find that our Comptonization model (BMC)
provides a good fit ($\chi^2$=268/266) with parameters $T=0.11 \pm 0.07$, 
keV and spectral index $\alpha=0.61\pm0.05$, typical a for low/hard state
accreting black hole. 
No additional components were required to obtain a reasonable spectral fit.
The inferred temperature $T$ and the normalization of the 
blackbody component $A_n$ allow us 
to estimate the BH mass for this source. The value we obtain 
is of order of $4\times 10^3$ solar masses 
(see Table 3). As previously noted in Titarchuk \& Shrader (2002) (see also 
Merloni, Fabian \& Ross 2002), typically, problems with the 
interpretation of X-ray spectral data are driven either by 
the application  of an additive, phenomenological model of  
the spectrum or by the models related to the determination 
of the inner disk temperature and
radius. One must bear in mind that the temperature of the 
inner disk radius is not an observable 
quantity, rather it is a inferred parameter of this particular disk  model.
The contribution  to the spectrum of the innermost region can be
significantly overestimated as a result of approximating the
inner boundary conditions.
A self-consistent analysis of the disk emission area, 
in the case when the entire disk is exposed to the
observer, reveals that it exceeds  the inner-most area 
by an  order of magnitude. In practical terms, this means that there 
is little  spectral information on this region.

There are interpretations (Mukai et al 2003,  Fabbiano et al, 2003) which 
invoke an optically thick outflow from 
Eddington-limit accretors, for which the observed luminosities are 
compatible with stellar-mass central obJects. Thus, one might pose 
the question: is it the two different assumptions about the origin of the
soft component that lead to the different 
masses? In fact, the spectrum by itself does not 
provide the information on geometrical configuration of the emission 
area (see Titarchuk \& Lyubarskij 1995). One needs timing characteristics 
(QPOs, time lags, the power spectrum as a function energy) in order to 
reveal the geometrical configuration.
In our analysis we were guided by the similarity of the high/soft states 
in ULX to BHXRBs in which the presence of the high frequency QPOs is 
well established . In this sense we assume the soft component originates 
in the disk but not in ``an optically thick outflow''. Recent findings 
of ``high frequency'' QPOs by Strohmayer \& Mushotzky (2003) provide a very 
strong basis for our assumption. 
Indeed, the assumption that the disk radiation is the 
origin of the soft component plus the self-consistent procedure for the mass 
evaluation and comparison of our results to other methods support the
validation of our approach.

One can argue on the fact that there are a number of perfectly 
normal stellar-mass X-ray binaries have luminosities overlapping the 
ULX regime; see for example section 4.2.4 of the recent comprehensive review of 
black hole observations by McClintock and Remillard (2003).  
In the McClintock \& Remillard  (2003) review, they note in particular,
three examples of 
over-luminous BHXRBs: 4U~1453-47, V404~Cyg and V4641 Sgr. 
We have included 4U 1453-47 and V4641~Sgr in our analysis,  however, 
the data used for V4641~Sgr were not 
obtained during the apparent 1999 super-Eddington event (those data were 
all obtained with RXTE in a scanning configuration, which along with 
possible saturation issues leads to analysis difficulties). Also,
since the spectrum is in , or transitioning to the 
low-hard state, we made the assumption that the inner-disk radius had
receded to ~17 $R_s$ (see ST99).  For 4U~1543-47,
the analysis presented is based on RXTE observations of June 21, 2003,
at which time the source luminosity was about $2\times10^{38} ergs/cm^2/s$ 
assuming a distance of 6.4 kpc. 
In both cases, our  results are consistent with those objects being stellar-mass 
black holes.  The 1989 outburst of V404 Cyg was covered by GINGA, however, the 
high-level data products available from the HEASARC cover only several 
epochs of the late decay stage, several months after the May/June 1989 
peak. More fundamentally, the peak luminosities exhibited by  
these sources are aperiodic intensity surges of relatively short duration, 
rather than the stable disk accretion scenarios that comprise 
the basis of our study.

Merloni, Fabian \& Ross (2000) stress  that  
disk inclination is a concern, as a rigorous treatment
must consider Doppler boosting effects, which can modify 
the emergent spectrum, particularly at high energies.
In principle this problem can be solved in the 
framework of a disk model with an assumption regarding distribution of 
energy release over the disk. In practice however, one would need to
know the disk equation of state, and then calculate the spectrum at each 
annulus, taking into account not only Doppler boosting effects, 
but the warping of the disk as well. In our view, there 
are simply too many uncertainties inherent in this approach. Rather, we 
argue that some effects of the Doppler boosting and disk warping can be 
implicitly present in the data through the spectral hardening factor $T_h$. 
Possibly it is not by chance $T_h=2.6$ is relatively high in all these 
BHs. The detailed study of this issue is beyond the scope of this 
paper. 

 We note that we have not considered the case of rapidly rotating black 
holes. Several objects among our sample, notably GRO~J1655-40 and GRS~1915+105 
have been cited as possible examples of extreme Kerr black holes on the basis of the
detection of high-frequency QPOs (e.g. Strohmayer 2001;  McClintock \ Remillard 2003 
and references therein), and XTE~1650-500 has been identified by Miller et 
al (2003) as a candidate rapidly-rotating object on the basis of spectroscopic
analysis. However, we note that there are concerns regarding the 
interpretation of the QPOs, notably the apparent synergy between BH, NS and WD systems 
(Mauche 2002), which argues against models invoking general relativistic
effects. Furthermore, alternative explanations have been put forth 
(e.g. Titarchuk \& Wood 2002, Titarchuk 2003).
Additionally, we find no evidence in our own spectral analysis for the 
presence excess emission in the 4-6 keV region, although we are working with 
the coarse RXTE spectral resolution. We would further note that alternative 
explanations of those features has recently emerged 
(Titarchuk, Kazanas \& Becker 2003). Thus,  while
we acknowledge that, if corroborated by further study, the presence of 
rapidly rotating objects could alter our conclusions for some cases, we suggest 
that the issue is unresolved.
 
Our mass-distribution results for the NLS1s, namely that the distribution is
shifted towards lower values relative to other  AGN, is consistent with the
hypothesis that they may represent an early, evolutionary stage of the
broader  AGN class, in which the central black hole is still undergoing growth
through interaction with bulge material. If this is the case, one might
speculate that a larger fraction of soft-excess  AGN may be seen with
increasing redshift. This is consistent at least with results on the spatial
variance statistics for the soft/hard subsamples of the deep surveys, 
but nonetheless remains speculative.

\vspace{0.2in}

\centerline{\bf{ ACKNOWLEDGMENTS}}
This work made use of the High-Energy Astrophysics Science Archive
Research Center at the NASA Goddard Space Flight Center, as well as the
NASA Extragalactic Database (NED) at the NASA Jet Propulsion Laboratory.    
We appreciate very interesting questions and suggestions of the referee.

\newpage

% insert tables

\begin{deluxetable}{rrrrrrr} 
\tablecolumns{7} 
\tablewidth{0pc} 
\tablecaption{Galactic BHXRBs} 
\tablehead{ 
\colhead{Source ID}  & \colhead{$\alpha$}  & \colhead{$kT$ (keV)}  & \colhead{$d_{kpc}$} 
  & \colhead{$M_i^{\tablenotemark{a}}$} 
}
\startdata 
GRS 1758-258  &  1.47$\pm$0.03  &  0.30$\pm$0.04  &  8.5  & 5.7$\pm$1.7 \\
GX 339-4  &  1.31$\pm$0.14  &  0.85$\pm$0.06  &  4  &  9.0$\pm$3.8 \\
XTEJ1550-564  &  1.55$\pm$0.03  &  0.86$\pm$0.02  &  5  &  9.4$\pm$2.1 \\
XTE J1650-500  &  1.21$\pm$0.03  &  0.61$\pm$0.08  &  5  &  10.6$\pm$4.0 \\
LMC X-1  &  1.63$\pm$0.03  &  0.79$\pm$0.06  &  55  &  11.2$\pm$1.5 \\
Nova Muscae  &  1.36$\pm$0.04  &  0.86$\pm$0.07  &  4  &  12.5$\pm$1.6 \\
GRS 1915+105  &  1.7$\pm$0.10  &  0.90$\pm$0.08  &  11  &  13.3$\pm$4.0 \\
4U 1630-47  &  1.43$\pm$0.02  &  0.95$\pm$0.02  &  10  &  7.4$\pm$1.5 \\
XTE J1859-224  &  1.74$\pm$0.02  &  0.82$\pm$0.08  &  11  &  12.3$\pm$1.7 \\
4U 1547-43  &  1.43$\pm$0.01  &  0.75$\pm$0.01  &  6.4  &  14.8$\pm$1.6 \\
EXO 1846-031  &  1.88$\pm$0.07  &  0.75$\pm$0.07  &  6  &  12$\pm$5 \\
XTE J1755-32  &  1.10$\pm$0.04  &  0.66$\pm$0.01  &  8  &  8.4$\pm$4 \\
GRS 1739-278  &  1.58$\pm$0.06  &  0.92$\pm$0.09  &  8  &  7.7$\pm$4 \\
V4641 Sgr  &  1.97$\pm$0.10  & 0.32$\pm$0.029 & 9.5  &  7.5$\pm$2   \\

\tablenotetext{a}{$M_i=M(\cos{i})^{1/2}$}
\enddata 
\end{deluxetable} 

\begin{deluxetable}{rrrrrrrr} 
\tablecolumns{7} 
\tablewidth{0pc} 
\tablecaption{Narrow Line Seyfert-1 Galaxies}
\tablehead{ 
\colhead{Source ID}
  & \colhead{$\alpha$}  & \colhead{$kT$ (keV)}  & \colhead{$z$} & \colhead{$log M_i
  ^{\tablenotemark{a}}$} }
\startdata 
Ark 564  &  1.48$\pm$0.01  &  0.16$\pm$0.01  &  0.024  &  6.31$\pm$0.02  \\
H0707-495  &  1.26$\pm$0.08  &  0.10$\pm$0.00  &  0.040  &  6.72$\pm$0.00  \\
IRAS17020 4544  &  1.18$\pm$0.03  &  0.22$\pm$0.01  &  0.060  &  6.01$\pm$0.04  \\
IRAS20181-224  &  1.28$\pm$0.13  &  0.25$\pm$0.03  &  0.185  &  5.96$\pm$0.08  \\
IZw1  &  1.19$\pm$0.05  &  0.19$\pm$0.02  &  0.061  &  5.99$\pm$0.08  \\
Mkn335  &  0.90$\pm$0.03  &  0.16$\pm$0.01  &  0.026  &  6.01$\pm$0.06  \\
Mkn42  &  0.82$\pm$0.11  &  0.20$\pm$0.02  &  0.024  &  5.31$\pm$0.07  \\
NAB 0205 024  &  1.17$\pm$0.04  &  0.16$\pm$0.01  &  0.155  &  6.62$\pm$0.05  \\
NGC4051  &  0.93$\pm$0.01  &  0.09$\pm$0.00  &  0.002  &  5.89$\pm$0.00  \\
PG1211 143  &  1.01$\pm$0.03  &  0.11$\pm$0.01  &  0.080  &  6.83$\pm$0.05  \\
RE J1034 39  &  1.14$\pm$0.11  &  0.13$\pm$0.01  &  0.042  &  6.26$\pm$0.04  \\
RX J0148-27  &  0.98$\pm$0.10  &  0.12$\pm$0.02  &  0.121  &  6.77$\pm$0.10  \\
Mkn110  &  0.83$\pm$0.09  &  0.11$\pm$0.01  &  0.036  &  6.79$\pm$0.03  \\
PKS 0558-504  &  1.02$\pm$0.06  &  0.26$\pm$0.03  &  0.137  &  6.79$\pm$0.05  \\
Mkn766  &  0.87$\pm$0.02  &  0.08$\pm$0.00  &  0.012  &  6.09$\pm$0.04  \\
Ton S180  &  1.34$\pm$0.03  &  0.17$\pm$0.00  &  0.062  &  6.38$\pm$0.02  \\
PHL 1181  &  1.3$\pm$0.7    & 0.086$\pm$0.027 &  0.190  &  7.01$\pm$0.3   \\

\enddata
\tablenotetext{a}{$M_i=M(\cos{i})^{1/2}$} 
\end{deluxetable} 

\begin{deluxetable}{rrrrrrrrrr} 
\tablecolumns{7} 
\tablewidth{0pc} 
\tablecaption{Ultra-Luminous non-Nuclear Extragalactic X-Ray Sources}
\tablehead{ 
\colhead{Source ID}
  & \colhead{$R.A.$}  & \colhead{$Dec$}  & \colhead{$kT$ (keV)}  & \colhead{$\alpha$}  
& \colhead{$d_{kpc}$} & \colhead{$log(M_i)^{\tablenotemark{a}}$} }
\startdata 
M33  &  01 33 51.1  &  +30 39 37  &  0.39$\pm$ 0.01  &  1.41$\pm$ 0.06  &  700  &  1.46$\pm$ 0.19  \\
NGC 1313  &  03 18 20.3  &  -66 29 11  &  0.28$\pm$ 0.06  &  1.31$\pm$ 0.02  &  3700  &  2.22$\pm$ 0.05 \\
NGC 5408  &  14 03 19.4  &  -42 22 58  &  0.11$\pm$ 0.01  &  1.20$\pm$ 0.10  &  4900  &  2.03$\pm$ 0.05 \\
NGC  253   &  00~47~17.6  &  -25~18~11  &  0.12 $\pm$ 0.03 &  1.1$\pm$0.24    &  3100  &   3.01$\pm$ 0.13 \\
NGC 1399~X-2 &  03~38~31.8& -35~26~04 &  0.102 $\pm$ 0.022 &  2.18$\pm$ 0.43    &  20000 &   3.54$\pm$ 0.16\\
NGC 1399~X-4 &  03~38~27.6& -35~26~48 &  0.15 $\pm$ 0.08 &  2.5$\pm$ 1.4    &  20000 &   3.34$\pm$ 0.22\\
IC 342~X-1 & 03~45~59  & 68~05~05 &  0.11 $\pm$ 0.01 &  0.62$\pm$ 0.05    & 4000 &   3.63$\pm$ 0.30\\
\enddata 
\tablenotetext{a}{$M_i=M(\cos{i})^{1/2}$}
\end{deluxetable} 
\newpage

\begin{figure}
\includegraphics[width=7in,height=6in,angle=0]{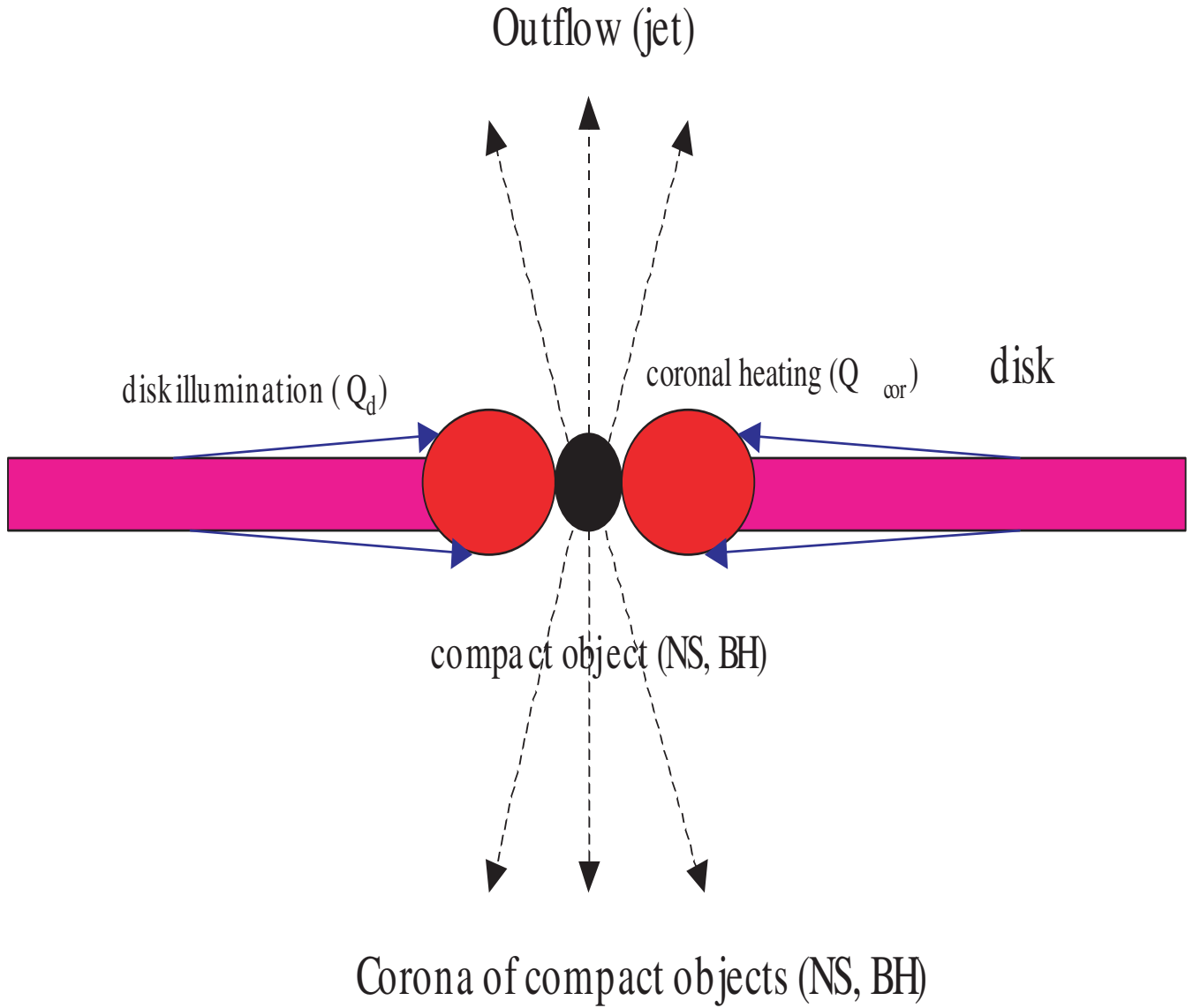}
\caption{
Artist's  conception of our model.
The low-energy blackbody X-ray photons are assumed  to come 
directly from the thin accretion disk. The higher energy photons 
that form the  power law tail of the
spectrum come from upscattering (acceleration) by energetic electrons depicted
in circular regions internal to the disk. 
Some fraction of the disk photons are viewed directly, while another 
fraction are seen after scattering in the coronal region.
}
\end{figure}
\begin{figure}
\epsscale{0.45}
\plotone{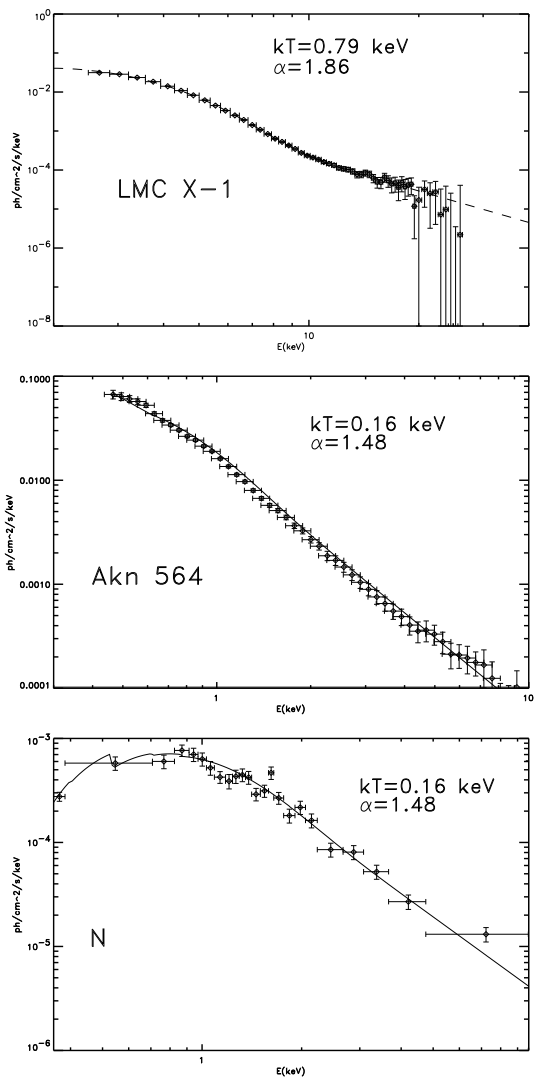}
% fig 1
\caption{
Representative spectral energy distributions for a BHXRB (LMC~X-1), 
a NLS1 (Akn~564), and an ULX (NGC~1313). Plotted are the data, corrected for 
the instrumental response (points), and the model as described in the text. Note 
that the basic power-law plus soft-excess form is a common feature, although
the characteristic black-body temperatures $kT$ differ ($\sim 1$ keV, $\sim 0.5$ keV,
 and $\sim0.2$ keV for LMC X-1, NLS1 and NGC 1313 respectively).
Note that $\alpha$ is the energy flux spectral index.}
\label{fig1}
\end{figure}
\newpage

\begin{figure}
\includegraphics[width=5in,height=5.5in,angle=0]{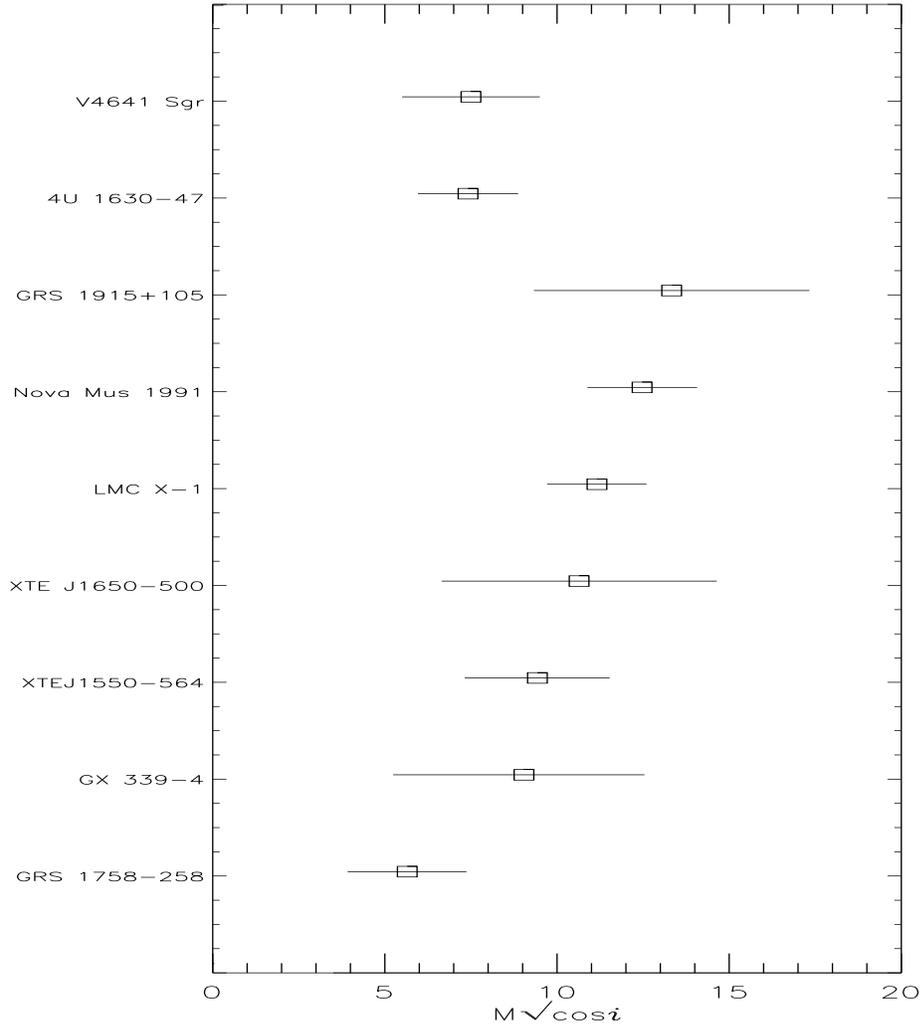}
\caption{The horizontal axis in this case is $M_i=M_{BH}\cos{i}$ in solar units, while the vertical axis
delineates between objects. Other than Nova Muscae 1991, and
possibly LMC X-1 (but see ShT99), our results are in reasonable agreement
with dynamical mass determinations where such measurements are
available.}
\end{figure}

%\begin{figure}
%\epsscale{0.55}
%\plotone{F2.eps}
% fig 2
%\caption{
% Distribution of our mass-determinations for Galactic BHXRBs..  The
%horizontal axis in this case is $M_{BH}$ in solar units, while the vertical axis
%delineates between objects. Other than Nova Muscae 1991, and
%possibly LMC X-1 (but see ShT99), our results are in reasonable agreement
%with dynamical mass determinations where such measurements are
%available.}
%\label{fig2}
%\end{figure}
\newpage

%\newpage
\begin{figure}
\includegraphics[width=5in,height=6.3in,angle=0]{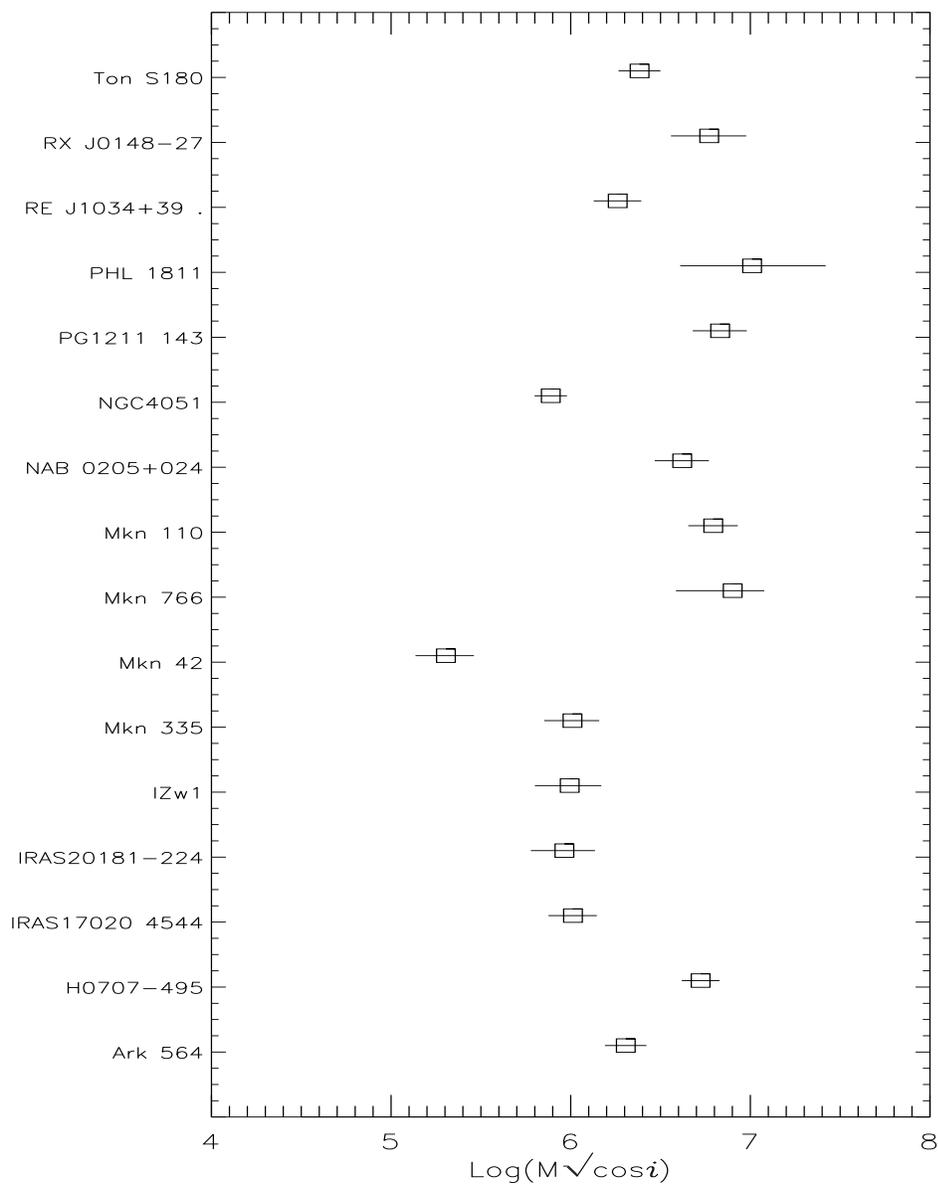}
\caption{Distribution of our mass-determination results for our NLS1s sample. The
horizontal axis in this case is $Log(M_i)$, while the vertical axis 
delineates between objects. $M_i=M_{BH}\cos{i}$ in solar units. 
The error-bars represent the uncertainties
associated with the color factor, which as noted in the text can vary over a wide
range for  AGN disks. }
\end{figure}

%\begin{figure}
%\epsscale{0.55}
%\plotone{f3.eps}
% fig 3
%\caption{
%Distribution of our mass-determination results for our NLS1s sample. The
%horizontal axis in this case is $Log(M_{BH})$, while the vertical axis 
%delineates between objects. The error-bars represent the uncertainties
%associated with the color factor, which as noted in the text can vary over a wide
%range for  GN disks. }
%\label{fig3}
%\end{figure}

\newpage
\begin{figure}
\includegraphics[width=5in,height=5.5in,angle=0]{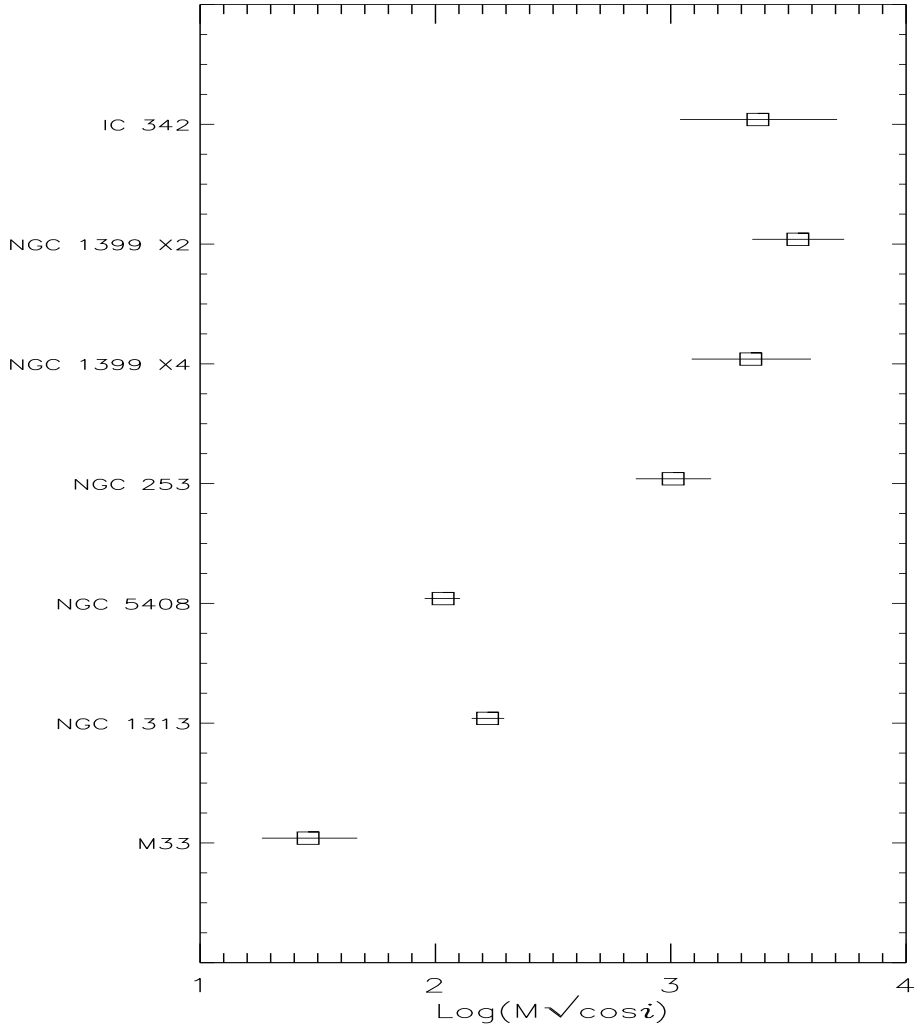}
\caption{Mass-determination results for three well known extragalactic over-luminous non-nuclear
X-ray source sample. The horizontal axis in this case is $Log(M_i)$, while the vertical axis simply
delineates between objects. $M_i=M_{BH}\cos{i}$ in solar units. 
The error-bars, again represent essentially the uncertainties
associated the color factors. }
\end{figure}

%\begin{figure}
%\epsscale{0.55}
%\plotone{f4.eps}
% fig 4
%\caption{
%Mass-determination results for three well known extragalactic over-luminous non-nuclear
%X-ray source sample. The horizontal axis in this case is $Log(M_i)$, while the vertical axis simply
%delineates between objects. $M_i=M_{BH}\cos{i}$ in solar units. 
%The error-bars, again represent essentially the uncertainties
%associated the color factors. }
%\label{fig4}
%\end{figure}

\newpage
\begin{figure}
\includegraphics[width=5in,height=6in,angle=0]{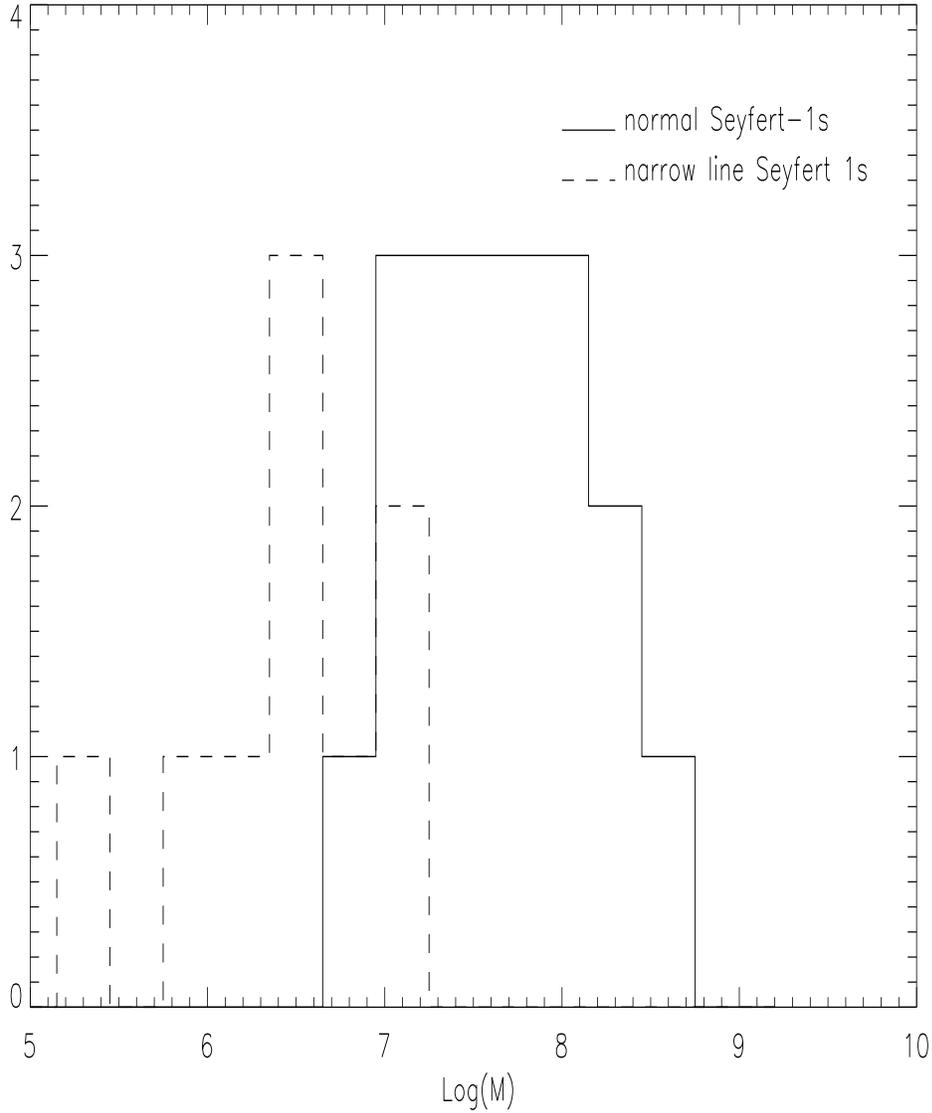}
\caption{Distribution of our NLS1 BH mass estimations. For comparison, a
sample of normal Seyfert-1 nuclear BH mass estimates from the literature
(Woo \& Urry, 2002) are shown as well. Our results suggest, in the absence
of some unforeseen systematic error in our analysis,  that the NLS1
population comprise a lower-mass Seyfert sub-population.}
\end{figure}

%\begin{figure}
%\epsscale{0.55}
%\plotone{f5.eps}
% fig 5
%
%\vspace{10pt}
%\caption{ Distribution of our NLS1 BH mass estimations. For comparison, a
%$sample of normal Seyfert-1 nuclear BH mass estimates from the literature
%(Woo \& Urry, 2002) are shown as well. Our results suggest, in the absence
%of some unforeseen systematic error in our analysis,  that the NLS1
%population comprise a lower-mass Seyfert sub-population.
% }
%\label{fig5}
%\end{figure}

\end{document}